\documentclass[12pt]{iopart}
\usepackage{graphicx}
\usepackage[usenames]{color}
\usepackage{amssymb}
\usepackage[dvips]{epsfig}

\begin{document}

\title[Bose-Einstein condensates in nonlinear lattices]{%
Quasi-one-dimensional Bose-Einstein condensates in nonlinear lattices}
\author{L. Salasnich$^{1}$ and B.A. Malomed$^{2}$}

\address{$^1$Dipartimento di Fisica ``Galileo Galilei''
and CNISM, Universit\`a di
Padova, Via Marzolo 8, 35131 Padova, Italy \\
$^{2}$Department of Physical Electronics, School of Electrical
Engineering, Faculty of Engineering, Tel Aviv University, Tel Aviv
69978, Israel}

\ead{luca.salasnich@unipd.it}

\begin{abstract}
We consider the three-dimensional (3D) mean-field model for the
Bose-Einstein condensate (BEC), with a 1D nonlinear lattice (NL),
which periodically changes the sign of the nonlinearity along the
axial direction, and the harmonic-oscillator trapping potential
applied in the transverse plane. The lattice can be created as an
optical or magnetic one, by means of available experimental
techniques. The objective is to identify stable 3D solitons
supported by the setting. Two methods are developed for this
purpose: The variational approximation, formulated in the framework
of the 3D Gross-Pitaevskii equation, and the 1D nonpolynomial
Schr\"{o}dinger equation (NPSE) in the axial direction, which allows
one to predict the collapse in the framework of the 1D description.
Results are summarized in the form of a stability region for the
solitons in the plane of the NL strength and wavenumber. Both
methods produce a similar form of the stability region. Unlike their
counterparts supported by the NL in the 1D model with the cubic
nonlinearity, kicked solitons of the NPSE cannot be set in motion,
but the kick may help to stabilize them against the collapse, by
causing the solitons to shed excess norm. A dynamical effect
specific to the NL is found in the form of freely propagating
small-amplitude wave packets emitted by perturbed solitons.
\end{abstract}

\maketitle

%\pacs{03.75.Lm, 05.45.Yv, 42.65.Tg}

\section{Introduction and the model}

The use of periodic potentials, induced by optical lattices, for steering
matter waves in Bose-Einstein condensates (BECs) is a vast research area, as
demonstrated in reviews \cite{ref2}-\cite{ref10}. An important aspect of
this topic is that the lattice potentials, balancing the cubic
nonlinearities induced by inter-atomic collisions in the BEC, help to create
and stabilize solitons. In particular, the lattices play a critical role in
stabilizing two-dimensional (2D) solitons against the collapse in the
condensate with intrinsic self-attraction \cite{2Dstabilization}.

Theoretical and experimental studies of the soliton dynamics in periodic
potentials were recently extended for \textit{nonlinear lattices }(NLs),
which may be induced by a spatially periodic modulation of the local
strength of\ the nonlinearity (the respective effective nonlinear potentials
are often called \textit{pseudopotentials} \cite{pseudo}). In BEC, the
spatial modulation can be implemented by means of the Feshbach resonance
controlled by properly patterned external fields (in the quasi-1D BEC, a
combination of linear and nonlinear lattices may also be induced by
periodically modulating the strength of the tight transverse confinement in
the axial direction \cite{Luca}). As well as their linear counterparts, NLs
have drawn much attention in connection to their potential for the creation
and control of matter-wave solitons in a number of different settings, see
original works \cite{HS}-\cite{2D} and review \cite{ref10}. While NLs
readily support stable 1D solitons \cite{HS}-\cite{1DKwok}, it has been
found difficult, albeit sometimes possible, to stabilize 2D solitons against
the collapse by means of the NL-induced pseudopotentials \cite{2D}. As
concerns the 1D solitons in models with the cubic nonlinearity, the
numerical analysis also reveals that they feature mobility in the presence
of NLs \cite{HS,mobility}.

Thus far, no example of stabilization of 3D solitons by NLs has been
reported. On the other hand, the creation of multidimensional solitons can
be facilitated by combinations of linear and nonlinear lattices, as shown,
in particular, in Ref. \cite{mixed}, where 2D solitons supported by crossed
1D lattices, one linear and one nonlinear, were reported. Still earlier, it
was demonstrated that the 1D linear-lattice potential, acting together with
periodic temporal modulations of the nonlinearity, which may be induced by
the Feshbach resonance controlled by a time-periodic external field, can
stabilize 3D solitons \cite{Michal}.

A natural possibility for the creation of 3D solitons is suggested by a
combination of the 1D NL with the usual harmonic-oscillator linear trapping
potential acting in the plane perpendicular to the NL axis (similar
settings, but with linear 1D lattices, were shown to be efficient in the
creation of 3D gap solitons in the BEC with the repulsive intrinsic
nonlinearity \cite{Canary}). This setting is the subject of the present
paper. We tackle it by means of two different methods, namely, the
variational approximation (VA)\ applied to the underlying 3D
Gross-Pitaevskii equation, and, in a more accurate form, the effective 1D
nonpolynomial Schr\"{o}dinger equation (NPSE), which was efficient in
description of many other settings dominated by the interplay of the tight
2D confinement and nonlinearity \cite{sala-npse}, \cite{Delgado}, \cite{Luca}%
, including the action of linear lattice potentials in the axial direction
\cite{we-lattice}.

Thus, we consider a dilute BEC of bosons with mass $m$ confined in the
transverse plane by the isotropic harmonic-oscillator potential with
frequency $\omega _{\perp }$, $V(x,y)={(1/2)}m\omega _{\bot }^{2}\left(
x^{2}+y^{2}\right) $. The corresponding 3D Gross-Pitaevskii equation is
rescaled by measuring time and coordinates in units of $\omega _{\perp }^{-1}
$ and the transverse-confinement radius, $a_{\bot }=\sqrt{\hbar /(m\omega
_{\bot })}$, respectively (hence the energy is measured in units of $\hbar
\omega _{\bot }$):
\begin{equation}
i{\frac{\partial \psi }{\partial t}}=\left[ -{\frac{1}{2}}\nabla ^{2}+{\frac{%
1}{2}}\left( x^{2}+y^{2}\right) +2\pi g(z)|\psi |^{2}\right] \psi ,
\label{3dgpe}
\end{equation}%
with the condensate's wave function normalized to unity,
\begin{equation}
\int |\psi (\mathbf{r},t)|^{2}d^{3}\mathbf{r}=1.  \label{1}
\end{equation}%
The interaction strength in Eq. (\ref{3dgpe}) is%
\begin{equation}
g(z)=2(N-1)a_{s}(z)/a_{\bot },  \label{g}
\end{equation}%
where $N$ is the number of atoms and $a_{s}$ the $z$-dependent \textit{s}%
-wave scattering length of the inter-atomic potential, the NL corresponding
to the periodic dependence,
\begin{equation}
g(z)=g_{0}+g_{1}\cos {(2kz)}.  \label{gdiz}
\end{equation}%
Below, we consider the most fundamental case of $g_{0}=0$. Placing the
center of the soliton at $z=0$, we assume $g_{1}<0$, to support the soliton
by the locally attractive nonlinearity.

The 1D periodic modulation of the local scattering length, implied by Eqs. (%
\ref{gdiz}) and (\ref{g}), can be implemented in an optical lattice,
produced by the interference of a pair of counterpropagating laser beams
controlling $a_{s}$ via the Feshbach resonance \cite{optics}. In that case,
the period of the resulting NL\ is limited by diffraction to $\gtrsim 1$ $%
\mathrm{\mu }$m. More often, the Feshbach resonance in experiments with BEC
is controlled by the magnetic field \cite{magnetic}. In that case, the 1D
periodic structure can be built as a magnetic lattice, imposed by a properly
designed set of ferromagnet films \cite{magn-latt}, with the respective
fabrication limit on the NL period also amounting to $\gtrsim 1$ $\mathrm{%
\mu }$m. Then, assuming that the trapping potential confines the transverse
size of the condensate, as usual, to the same order of magnitude (a few
microns), one may conclude that the solitons are built of several thousand
atoms \cite{Randy}.

The rest of the paper is organized as follows. The direct VA is developed in
Section II, and the effective 1D NPSE is derived in Section III. In both
cases, we find the stability region for solitons in the plane of $\left(
k,\left\vert g_{1}\right\vert \right) $. The mobility of the solitons is
tested in Section III by means of direct simulations of the evolution of
axially kicked solitons, in the framework of the NPSE (it is concluded that
the solitons are not mobile in the present setting; however, the kick may
effectively stabilize solitons against the collapse). The paper is concluded
by Section IV.

\section{The variational approximation}

Our first aim is to apply the VA to Eq. (\ref{3dgpe}), following the lines
of Ref. \cite{sala-time}. To this end, we notice\ that Eq. (\ref{3dgpe}) can
be derived from the Lagrangian density,
\begin{equation}
\mathcal{L}={\frac{i}{2}}\big(\psi ^{\ast }\frac{\partial \psi }{\partial t}%
-\psi \frac{\partial \psi ^{\ast }}{\partial t}\big)-{\frac{1}{2}}|\nabla
\psi |^{2}-{\frac{1}{2}}\left( x^{2}+y^{2}\right) |\psi |^{2}-\pi g(z)|\psi
|^{4},  \label{lagrangian}
\end{equation}%
and make use of a time-dependent Gaussian ansatz,
\begin{equation}
\psi \left( \mathbf{r},t\right) ={\frac{\exp \left\{ -\frac{1}{2}\left[
\frac{r_{\bot }^{2}}{\sigma _{\bot }^{2}(t)}+\frac{z^{2}}{\sigma _{\parallel
}^{2}(t)}\right] +i\beta _{\bot }(t)r_{\bot }^{2}+i\beta _{\parallel
}(t)z^{2}\right\} }{\pi ^{3/4}{\sigma _{\bot }}(t)\sqrt{{\sigma }_{\parallel
}(t)}}}\;,  \label{ansatz}
\end{equation}%
where $r_{\bot }^{2}\equiv x^{2}+y^{2},$ and $\sigma _{\bot }(t)$, $\sigma
_{\parallel }(t)$ and $\beta _{\bot }(t)$, $\beta _{\parallel }(t)$ are
time-dependent variational parameters. This wave function is an exact one
for non-interacting bosons ($g=0$) in the harmonic trap.

Inserting the ansatz into Lagrangian density (\ref{lagrangian}) and
performing the spatial integration, we arrive at the effective Lagrangian,
\begin{eqnarray}
L &=&-{\frac{1}{2}}\Big[\big(2{\dot{\beta}_{\bot }}\sigma _{\bot }^{2}+{%
\frac{1}{\sigma _{\bot }^{2}}}+4\sigma _{\bot }^{2}\beta _{\bot }^{2}+\sigma
_{\bot }^{2}\big) \\
&+&\big({\dot{\beta}_{\parallel }}\sigma _{\parallel }^{2}+{\frac{1}{2\sigma
_{\parallel }^{2}}}+2\sigma _{\parallel }^{2}\beta _{\parallel }^{2}\big)+{%
\frac{g_{0}+g_{1}e^{-k^{2}\sigma _{\parallel }^{2}/2}}{\sqrt{2\pi }\ \sigma
_{\bot }^{2}\sigma _{\parallel }}}\Big],\;
\end{eqnarray}%
with the overdot standing for time derivatives. The respective
Euler-Lagrange equations take the form of
\begin{eqnarray}
\beta _{\bot } &=&-{\frac{{\dot{\sigma}_{\bot }}}{2\sigma _{\bot }}}\;,
\label{uffa1a} \\
\beta _{\parallel } &=&-{\frac{{\dot{\sigma}_{\parallel }}}{2\sigma
_{\parallel }}}\;,  \label{uffa1b} \\
{\ddot{\sigma}_{\bot }}+\sigma _{\bot } &=&{\frac{1}{\sigma _{\bot }^{3}}}+{%
\frac{g_{0}+g_{1}e^{-k^{2}\sigma _{\parallel }^{2}/2}}{\sqrt{2\pi }\ \sigma
_{\bot }^{3}\sigma _{\parallel }}}\;,  \label{uffa2a} \\
{\ddot{\sigma}_{\parallel }} &=&{\frac{1}{\sigma _{\parallel }^{3}}}+{\frac{%
g_{0}+g_{1}e^{-k^{2}\sigma _{\parallel }^{2}/2}(1+k^{2}\sigma _{\parallel
}^{2})}{\sqrt{2\pi }\ \sigma _{\bot }^{2}\sigma _{\parallel }^{2}}}\;.
\label{uffa2b}
\end{eqnarray}%
Equations (\ref{uffa1a}) and (\ref{uffa1b}) show that, as usual, chirps $%
\beta _{\bot }$ and $\beta _{\parallel }$ are determined by the time
dependence of $\sigma _{\bot }$ and $\sigma _{\parallel }$, while Eqs. (\ref%
{uffa2a}) and (\ref{uffa2b}) correspond to the equations of motion of a
mechanical system with two degrees of freedom, whose energy is
\[
E={\frac{1}{2}}\dot{\sigma}_{\bot }^{2}+{\frac{1}{4}}\dot{\sigma}_{\parallel
}^{3}+U(\sigma _{\bot },\sigma _{\parallel }),
\]%
\begin{equation}
U(\sigma _{\bot },\sigma _{\parallel })={\frac{1}{2}}\sigma _{\bot }^{2}+{%
\frac{1}{2\sigma _{\bot }^{2}}}+{\frac{1}{4\sigma _{\parallel }^{2}}}+{\frac{%
g_{0}+g_{1}e^{-k^{2}\sigma _{\parallel }^{2}/2}}{2\sqrt{2\pi }\ \sigma
_{\bot }^{2}\sigma _{\parallel }}}\;.  \label{U}
\end{equation}

Next, we look for stationary configurations corresponding to minima of
potential energy (\ref{U}), by demanding $\partial U/\partial \sigma _{\bot
}=\partial U/\partial \sigma _{\parallel }=0$, which yields
\begin{eqnarray}
\sigma _{\bot } &=&{\frac{1}{\sigma _{\bot }^{3}}}+{\frac{%
g_{0}+g_{1}e^{-k^{2}\sigma _{\parallel }^{2}/2}}{\sqrt{2\pi }\ \sigma _{\bot
}^{3}\sigma _{\parallel }}}\;,  \label{uffa2a-s} \\
0 &=&{\frac{1}{\sigma _{\parallel }^{3}}}+{\frac{g_{0}+g_{1}e^{-k^{2}\sigma
_{\parallel }^{2}/2}(1+k^{2}\sigma _{\parallel }^{2})}{\sqrt{2\pi }\ \sigma
_{\bot }^{2}\sigma _{\parallel }^{2}}}\;.  \label{uffa2b-s}
\end{eqnarray}%
These equations, which are fixed points of Eqs. (\ref{uffa2a}) and (\ref%
{uffa2b}), with ${\ddot{\sigma}_{\bot }}={\ddot{\sigma}_{\parallel }}=0$,
can be solved numerically. The solutions provide for a minimum of the energy
under the necessary condition that the Gaussian curvature $K_{G}$ of energy
surface $U(\sigma _{\bot },\sigma _{\parallel })$ is positive, i.e.,
\begin{equation}
K_{G}\equiv {\frac{\partial ^{2}U}{\partial \sigma _{\bot }^{2}}}{\frac{%
\partial ^{2}U}{\partial \sigma _{\parallel }^{2}}}-\left( {\frac{\partial
^{2}U}{\partial \sigma _{\bot }\partial \sigma _{\parallel }}}\right)
^{2}>0\;.  \label{curvature}
\end{equation}

Further, low-energy excitations of the condensate around the stationary
state are represented by small oscillations of variables $\sigma _{\bot }$
and $\sigma _{\parallel }$ around the equilibrium point defined by Eqs. (\ref%
{uffa2a-s}) and (\ref{uffa2b-s}). The calculation of the corresponding
normal-mode frequencies, $\Omega $, is thus reduced to finding eigenvalues
of the respective Hessian matrix,
\begin{equation}
\Lambda =\left(
\begin{array}{cc}
\partial ^{2}U/\partial \sigma _{\bot }^{2} & \partial ^{2}U/\partial \sigma
_{\bot }\partial \sigma _{\parallel } \\
\partial ^{2}U/\partial \sigma _{\parallel }\partial \sigma _{\bot } &
\partial ^{2}U/\partial \sigma _{\parallel }^{2}%
\end{array}%
\right) \;,  \label{hessian}
\end{equation}%
while the associated mass matrix is
\[
M=\left(
\begin{array}{cc}
1 & 0 \\
0 & {\frac{1}{2}}%
\end{array}%
\right) \;.
\]%
Then, the eigenfrequencies are found as the solutions of equation
\begin{equation}
\det \left( \Lambda -\Omega ^{2}M\right) =0\;.  \label{Omega}
\end{equation}

\begin{figure}[tbp]
\begin{center}
{\includegraphics[width=9.cm,clip]{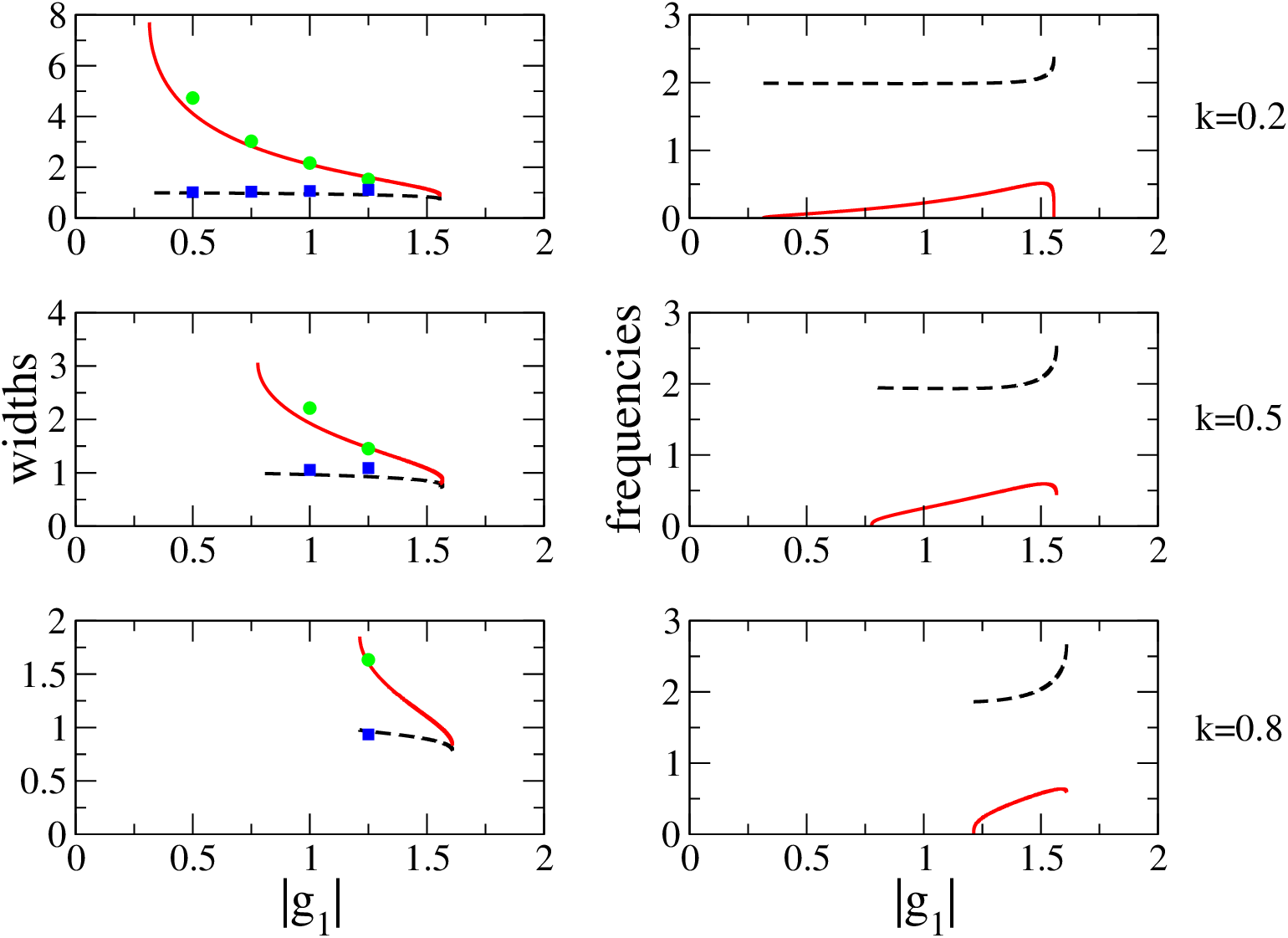}}
\end{center}
\caption{(Color online) Left panels: transverse and axial widths, $\protect%
\sigma _{\bot }$ and $\protect\sigma _{\parallel }$ (dashed and solid lines,
respectively) of stable bright solitons, as functions of the NL strength, $%
|g_{1}|$, with $g_{0}=0$ and $g_{1}<0$, obtained from the variational
approximation. Filled circles and squares depict the widths obtained from a
numerical solution of the NPSE. Right panels: two eigenfrequencies $\Omega
_{1}$ and $\Omega _{2}$ (dashed and solid lines) of collective excitations
around the stable bright soliton, vs. $|g_{1}|$, as obtained from the
variational approximation. The results are shown for three values of the NL
wavenumber: $k=0.2$ (upper panels), $k=0.5$ (middle panels), and $k=0.8$
(lower panels).}
\label{fig1}
\end{figure}

Widths $\sigma _{\bot }$ and $\sigma _{\parallel }$ of stable solitons, as
found from the numerical solution of Eqs. (\ref{uffa2a-s}) and (\ref%
{uffa2b-s}), are plotted in the left panels of Fig. \ref{fig1} as functions
of the NL strength parameter, $|g_{1}|$ [recall we set $g_{0}=0$ and $g_{1}<0
$ in Eq. (\ref{gdiz})], along with the widths obtained by solving
numerically the 1D NPSE (see below). The comparison demonstrates close
proximity of the predictions of the VA to the results produced by the NPSE.
In the right panels of Fig. \ref{fig1}, we plot eigenfrequencies $\Omega _{1}
$ and $\Omega _{2}$ of excitations around the bright soliton, as found from
Eq. (\ref{Omega}).

\begin{figure}[tbp]
\begin{center}
{\includegraphics[width=7.5cm,clip]{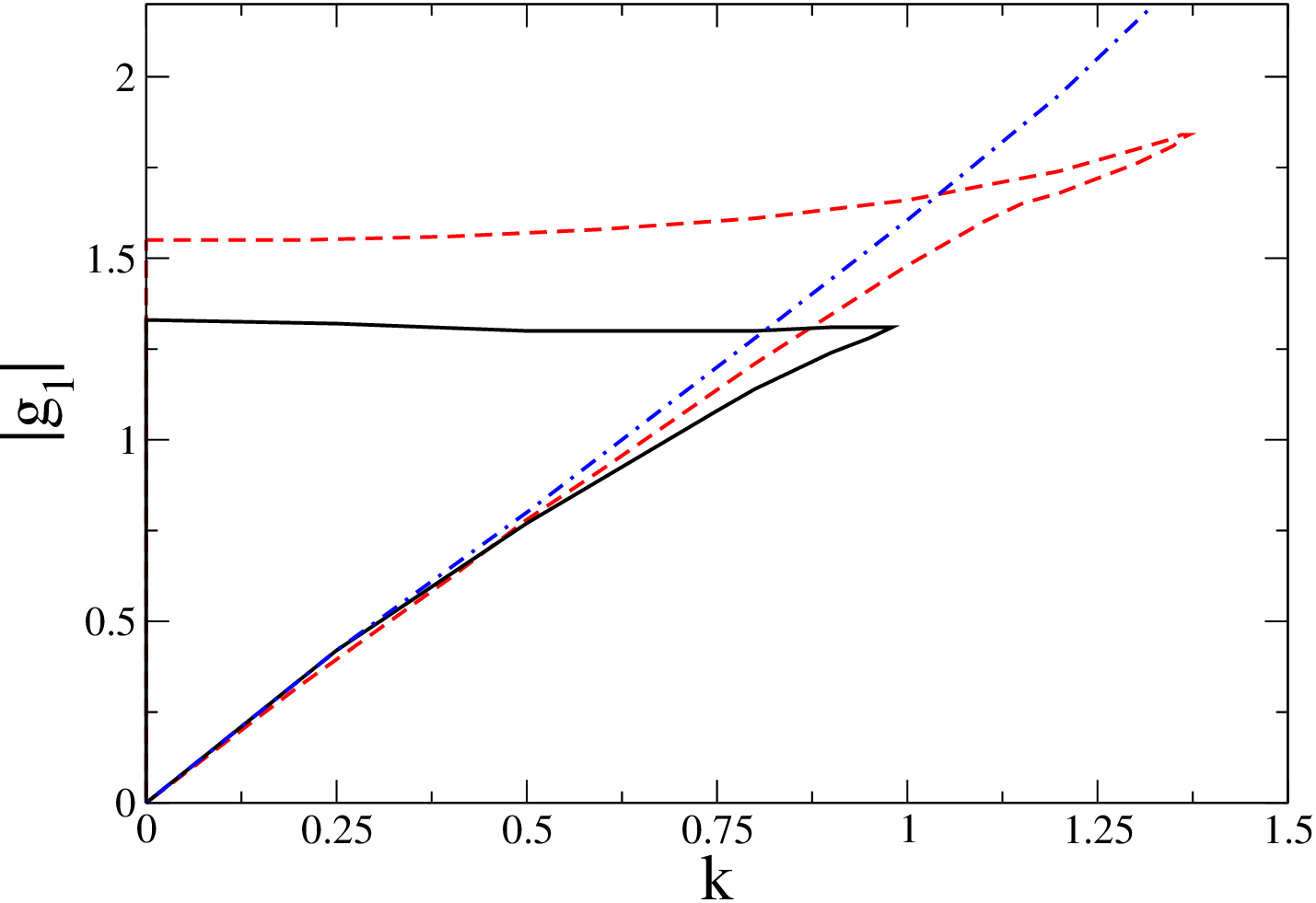}}
\end{center}
\caption{(Color online) The stability diagram for the solitons in the plane
of wavenumber $k$ and strength $|g_{1}|$ of the NL (with $g_{0}=0$). The
solitons are stable between the dashed lines, according to the variational
approximation, and between the solid lines, according to the NPSE. The
dot-dashed line is the lower bound predicted by the one-dimensional cubic
Gross-Pitaevskii equation.}
\label{fig2}
\end{figure}

In Fig. \ref{fig2} we plot the stability diagram in the plane of $(k,|g_{1}|)
$ for the bright solitons trapped in the NL with wavenumber $k$ and strength
$|g_{1}|$ (again, with $g_{0}=0$). The stability region, defined as that
where solutions of Eqs. (\ref{uffa2a-s}) and (\ref{uffa2b-s}) yield energy
minima, is bounded by the dashed lines (strictly speaking, the bright
solitons are only metastable,\ because for $g_{0}=0$ and $g_{1}<0$ the true
ground state is the collapsed one, with potential energy $U=-\infty $).
Below the lower dashed line, the soliton is subject to spreading along
longitudinal axis $z$ (which may be considered as a manifestation of the
delocalization transition, which was earlier studied in linear lattices \cite%
{BBB}). Above the upper dashed line, the soliton is destroyed by the
collapse. Very close to the upper dashed line, the VA predicts a
bistability, i.e., coexistence of two stable solitons at the same values of $%
k$ and $|g_{1}|$ (in Fig. \ref{fig1}, only the solitons with the lowest
energy are shown in the case of the bistability). In Fig. \ref{fig2} we also
plot the respective stability region (between the solid lines) as obtained
from the numerical solution of the NPSE (see below). The shrinkage and
disappearance of the stability region at large values of $k$ is quite
natural, as in that case the rapidly oscillating nonlinearity in Eq. (\ref%
{gdiz}) tends to average itself to zero.

The prediction of the usual 1D cubic Gross-Pitaevskii equation is also
plotted in Fig. \ref{fig2}. In that case, there is only one stability
boundary (the dot-dashed line), as the 1D equation with the cubic
nonlinearity does not predict the collapse. Accordingly, the above argument
concerning the disappearance of the stability region at large $k$ does not
apply to the cubic equation, because, in the limit of the small NL\ period, $%
\pi /k,$ the soliton may compress itself into a single potential well, and
this trend will not be aborted by the onset of the collapse.

\section{The nonpolynomial Schr\"{o}dinger equation (NPSE)}

\subsection{The derivation and imaginary-time evolution}

A more accurate description of the solitons is provided by the NPSE, which
can be derived by means of the semi-variational approach from the full 3D
equation (\ref{3dgpe}), using the method developed in Ref. \cite{sala-npse}.
To this end, we adopt the following ansatz, which, unlike the above one (\ref%
{ansatz}), contains arbitrary functions of the longitudinal coordinate, $%
\sigma (z,t)$ and $f(z,t)$, accounting for the transverse width and axial
wave function of the condensate:
\begin{equation}
\psi (\mathbf{r},t)={\frac{1}{\sqrt{\pi }\sigma (z,t)}}\exp {\left[ -{\frac{%
x^{2}+y^{2}}{2\left( \sigma (z,t)\right) ^{2}}}\right] }\,f(z,t).
\label{psi}
\end{equation}%
Note that, as follows from Eqs. (\ref{psi}) and (\ref{1}), the norm of the
axial wave function is also $1$:%
\begin{equation}
N_{\mathrm{1D}}\equiv \int_{-\infty }^{+\infty }\left\vert f(z)\right\vert
^{2}dz=1.  \label{N1D}
\end{equation}%
Substituting ansatz (\ref{psi}) into Lagrangian density (\ref{lagrangian}),
performing the integration over $x$ and $y$, and omitting spatial
derivatives of the transverse width (this corresponds to the adiabatic
approximation, which is known to produce accurate results in other settings
\cite{sala-npse}), we derive the respective Lagrangian density,
\begin{equation}
\bar{\mathcal{L}}={\frac{i}{2}}\big(f^{\ast }\frac{\partial f}{\partial t}-f%
\frac{\partial f^{\ast }}{\partial t}\big)-\frac{1}{2}\left\vert {\frac{%
\partial f}{\partial z}}\right\vert ^{2}-\frac{1}{2}\left( \frac{1}{\sigma
^{2}}+\sigma ^{2}\right) |f|^{2}-\frac{1}{2}g(z)\frac{|f|^{4}}{\sigma ^{2}}%
\;.  \label{effective}
\end{equation}%
Varying it with respect to $f^{\ast }(z,t)$ and $\sigma (z,t)$ gives rise to
the system of Euler-Lagrange equations:
\begin{eqnarray}
i\frac{\partial f}{\partial t} &=&\Big[-\frac{1}{2}\frac{\partial ^{2}}{%
\partial z^{2}}+\frac{1}{2}\left( \frac{1}{\sigma ^{2}}+\sigma ^{2}\right)
+g(z)\frac{|f|^{2}}{\sigma ^{2}}\Big]f\;,  \label{d-npse} \\
\sigma ^{4} &=&1+g(z)|f|^{2}\;,  \label{sigma}
\end{eqnarray}%
Inserting Eq. (\ref{sigma}) into Eq. (\ref{d-npse}), we obtain a closed-form
equation for the axial wave function, which is tantamount to the NPSE
derived in Ref. \cite{sala-npse}, but with the $z$-dependent interaction
strength, $g(z)$:
\begin{equation}
i\frac{\partial f}{\partial t}=\Big[-\frac{1}{2}\frac{\partial ^{2}}{%
\partial z^{2}}+{\frac{1+{(3/2)}g(z)|f|^{2}}{\sqrt{1+g(z)|f|^{2}}}}\Big]f\;.
\label{1dnpse}
\end{equation}

In the weak-coupling regime, i.e., $\left\vert g(z)\right\vert
|f(z,t)|^{2}\ll 1$, one can expand the nonpolynomial term in Eq. (\ref%
{1dnpse}), arriving at the cubic-quintic nonlinear Schr\"{o}dinger (with
term $1$ representing here the transverse ground-state energy),
\begin{equation}
i\frac{\partial f}{\partial t}=\Big[-\frac{1}{2}\frac{\partial ^{2}}{%
\partial z^{2}}+1+g(z)|f|^{2}+{\frac{3}{8}}g(z)^{2}|f|^{4}\Big]f\;.
%\label{weak-1dnpse}
\end{equation}%
The cubic-quintic nonlinearity for the tightly confined BEC was also derived
by means of different approaches \cite{CQ}. On the other hand, in the
strong-coupling regime, $g(z)|f(z,t)|^{2}\gg 1$ (which is relevant only for
the repulsive sign of the nonlinearity, $g>0$), the NPSE amounts to the
nonlinear Schr\"{o}dinger equation with the quadratic nonlinearity (see,
e.g., Ref. \cite{Mexico}):
\begin{equation}
i\frac{\partial f}{\partial t}=\Big[-\frac{1}{2}\frac{\partial ^{2}}{%
\partial z^{2}}+{\frac{3}{2}}\sqrt{g(z)}|f|\Big]f\;.  \label{strong-1dnpse}
\end{equation}

Here we aim to analyze bright solitons within the framework of Eq. (\ref%
{1dnpse}) with $g(z)$ given by Eq. (\ref{gdiz}), with $g_{0}=0$ and $g_{1}<0$%
, as said above. Results were obtained from numerical solutions based the
finite-difference Crank-Nicolson predictor-corrector algorithm \cite%
{sala-numerics}.

\begin{figure}[tbp]
\begin{center}
{\includegraphics[width=7.cm,clip]{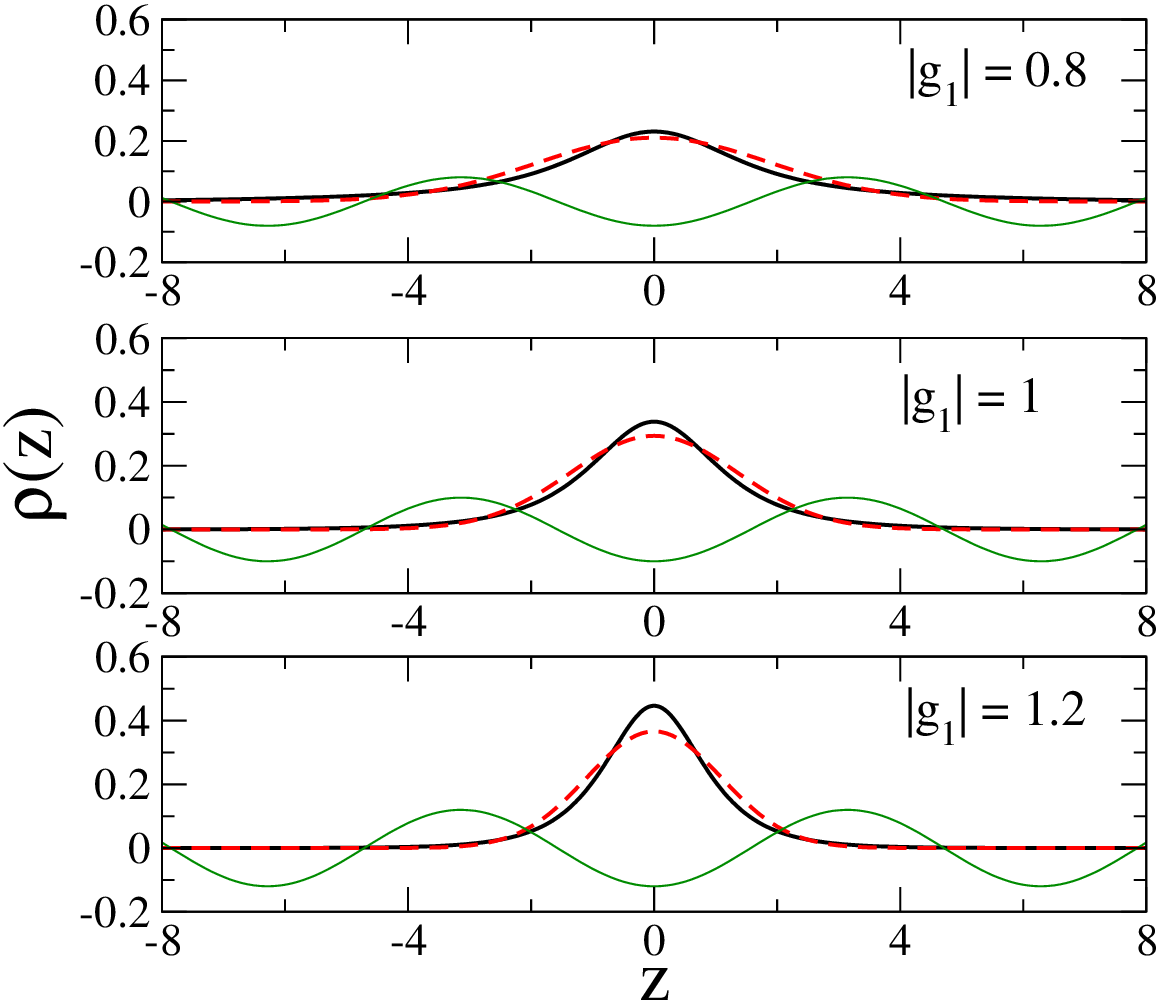}}
\end{center}
\caption{(Color online) Typical examples of axial density $\protect\rho %
(z)\equiv \left\vert f(z)\right\vert $ of stable solitons. The solid
and dashed lines display the results produced by the NPSE and
variational approximation based on ansatz (\protect\ref{ansatz}),
respectively, for three different values of the interaction strength
$|g_{1}|$, fixing $g_{0}=0 $ and $k=0.5$. Here and in Figs.
\protect\ref{fig4}, \protect\ref{fig5}, and \protect\ref{fig6}, the
green sinusoidal line represents the periodic
modulation function of the local nonlinearity defined in Eq. (\protect\ref%
{gdiz}).}
\label{fig3}
\end{figure}

First, by simulating the NPSE in imaginary time, we study the formation of
bright solitons. In particular, at $g_{1}=-0.4$, the soliton does not
self-trap, slowly degenerating towards a uniform configuration along axial
direction $z$. Instead, at $g_{1}=-1$ the bright soliton self-traps quickly,
representing the ground state of Eq. (\ref{1dnpse}). In Fig. \ref{fig3} we
plot typical examples of the axial density, $\rho (z)$, of the so obtained
stable bright soliton trapped in the NL, comparing the NPSE results to those
predicted by Gaussian ansatz (\ref{ansatz}) (the stability of the solitons
was verified by real-time simulations, see below). The figure shows that the
stable solitons are localized around one potential minimum of the NL (at $x=0
$). The NPSE profiles are quite close to their variational counterparts, see
also the left panels in Fig. \ref{fig1}. The main quantitative, although not
very large, difference between the VA and NPSE is the prediction of the
critical strength for the onset of the collapse, as seen in Fig. \ref{fig2}.

\section{Real-time dynamics}

\subsection{Stability of the solitons}

\begin{figure}[tbp]
\begin{center}
{\includegraphics[width=7.cm,clip]{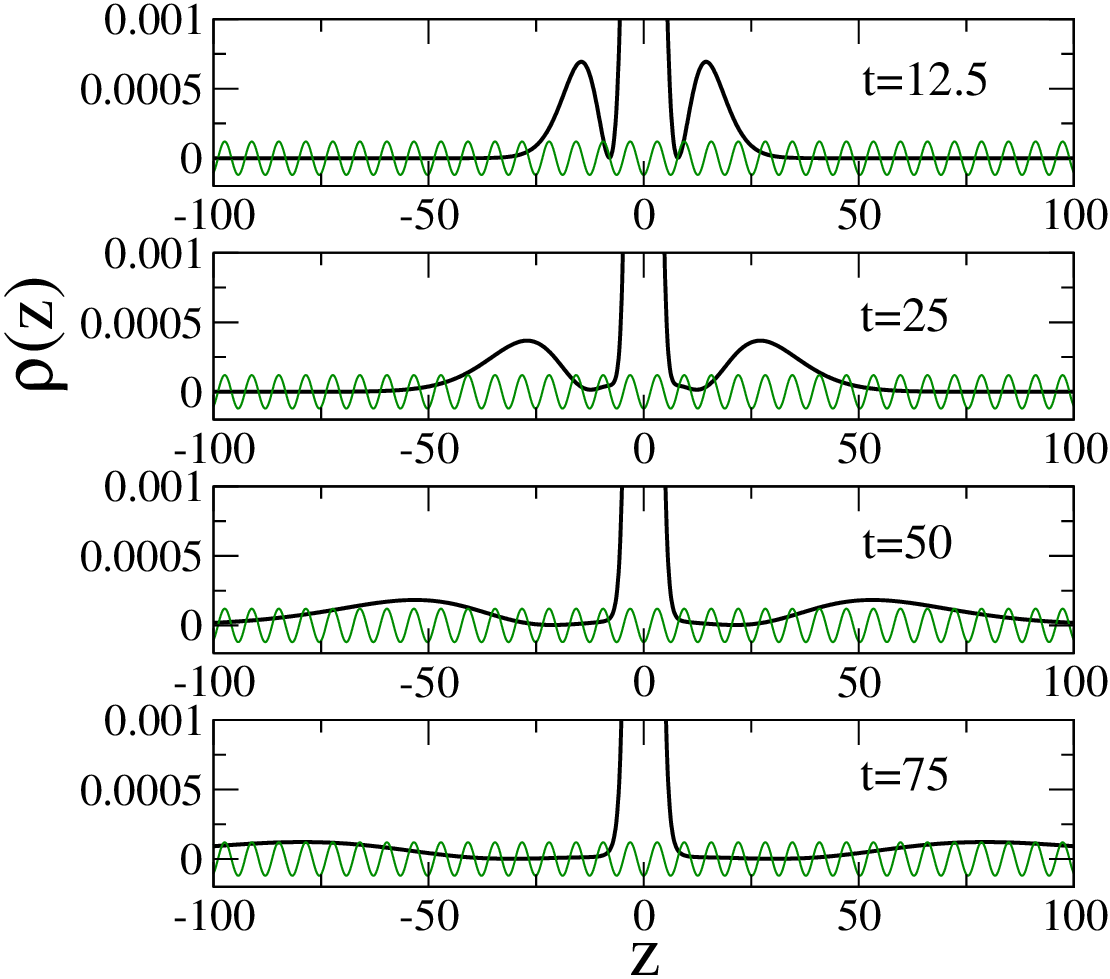}}
\end{center}
\caption{(Color online) Dynamics of a Gaussian wave packet with the initial
shape close to the ground-state soliton. The axial density, $\protect\rho (z)
$, is displayed at different values of real time $t$, as obtained from
simulations of Eq. (\protect\ref{1dnpse}). At $t=0$ (the initial condition,
not shown here), there is only the Gaussian centered at $z=0$, with axial
width $1.7$. The parameters are $g_{0}=0$, $g_{1}=-1.2$, and $k=0.5$.}
\label{fig4}
\end{figure}

The next step is the study of the real-time dynamics of the quasi-1D BEC
trapped in the NL. First of all, simulating NPSE (\ref{1dnpse}) in real
time, we have found that all the existing solitons are stable, with
initially perturbed wave functions featuring small oscillations around the
solitonic configurations. As shown in Fig. \ref{fig4}, an interesting
dynamical feature is observed if the initial wave function is the Gaussian
with a width close to that of the soliton: expulsion of two small waves from
the Gaussian peak (which is represented by the column centered at $z=0$ in
Fig. \ref{fig4}). The emitted waves rapidly move in opposite directions,
while the remaining central peak relaxes into a stationary soliton. This
effect is interesting because the same is not observed in linear lattices,
which would readily trap the radiation ``garbage" emitted by the central
peak, while the NL is not felt by the small-amplitude waves, hence they may
escape freely.

\subsection{Immobility of the trapped solitons}

The mobility of solitons trapped in the NL can be tested by applying a kick
to initially quiescent solitons \cite{HS}. For this purpose, Eq. (\ref%
{1dnpse}) was simulated with initial condition $f(z,t=0)=f_{\mathrm{sol}%
}(z)\ e^{ivz}$, where $v$ is the magnitude of kick, i.e., the initial
velocity imparted to soliton $f_{\mathrm{sol}}(z)$, which was produced by
means of the imaginary-time simulations of the same NPSE. To present a
typical result, we fix $g_{0}=0$, $g_{1}=-1.2$, and $k=0.5$, and perform
real-time simulations at increasing values of $v$. As shown in Fig. \ref%
{fig5}, at $v=0.4$ we observe ejection of small-amplitude waves from the
soliton (cf. Fig. \ref{fig4}), while the central peak remains trapped at the
initial position, relaxing back into a stationary soliton, with a somewhat
smaller value of the norm.

\begin{figure}[tbp]
\begin{center}
{\includegraphics[width=7.cm,clip]{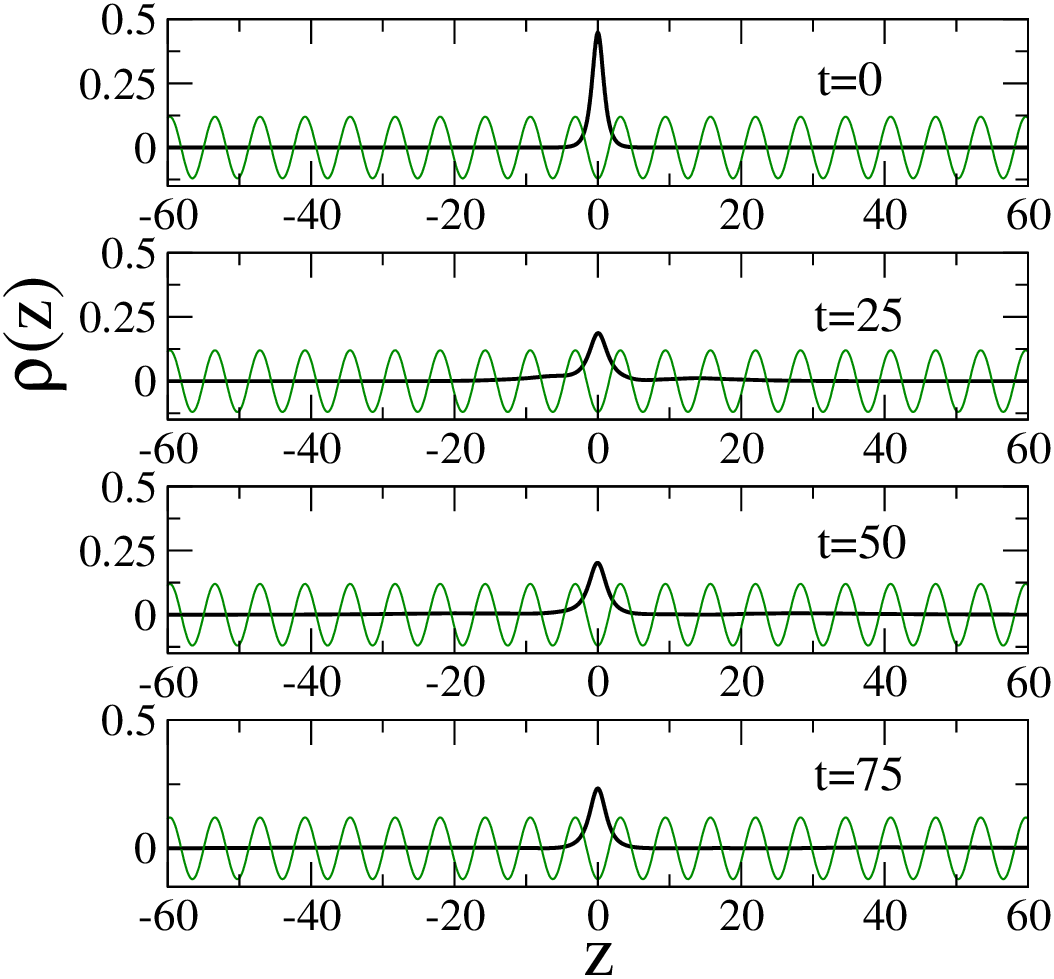}}
\end{center}
\caption{(Color online) The evolution of the kicked soliton with initial
velocity $v=0.4$. Axial density $\protect\rho (z)$ is plotted at different
values of real time $t$, as obtained from simulations of Eq. (\protect\ref%
{1dnpse}). Here, $g_{0}=0$, $g_{1}=-1.2$, and $k=0.5$ are fixed.}
\label{fig5}
\end{figure}

\begin{figure}[tbp]
\begin{center}
{\includegraphics[width=7.cm,clip]{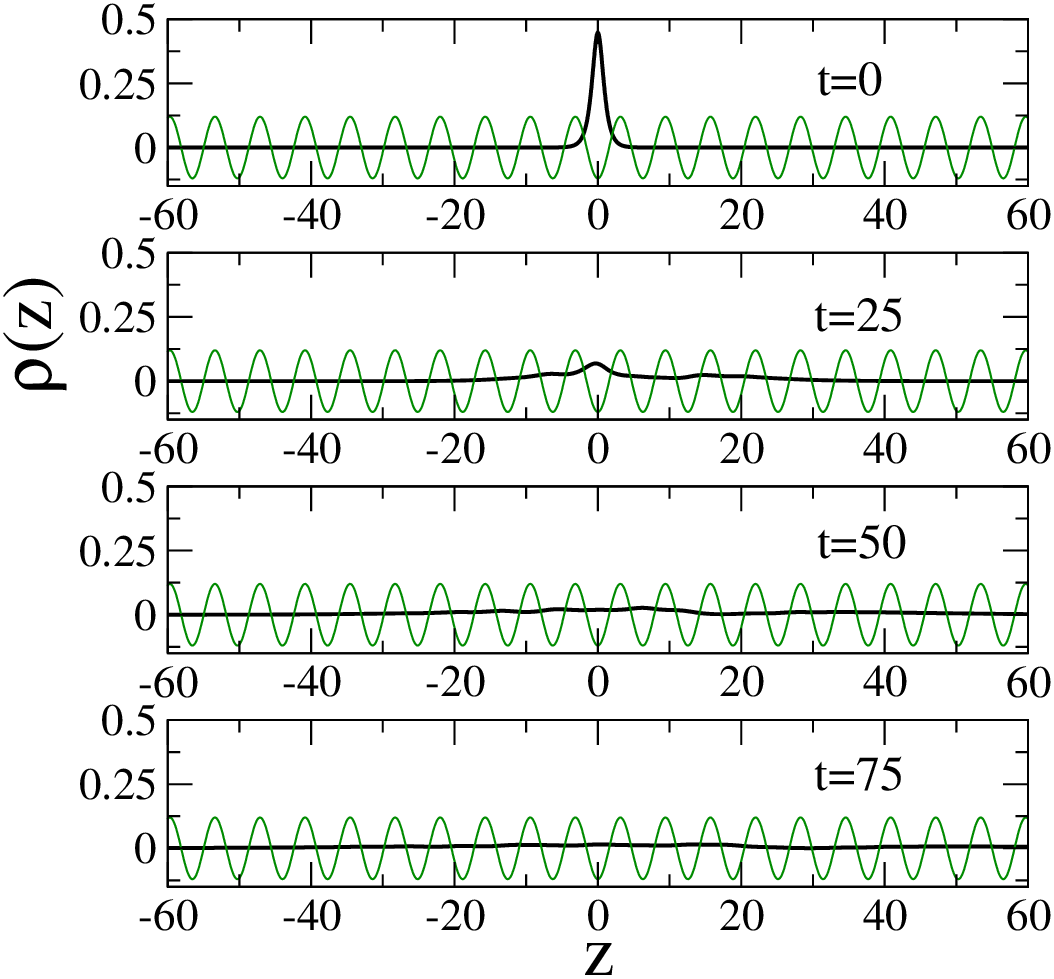}}
\end{center}
\caption{(Color online) The same as in Fig. \protect\ref{fig5}, but
for the soliton kicked with initial velocity $v=0.6$.} \label{fig6}
\end{figure}

With the increase of $v$, the amplitude and the velocity of the ejected
waves increases, but the remaining soliton stays put. This is a noteworthy
difference from the soliton dynamics in 1D NLs with the cubic nonlinearity,
where the soliton may be set in motion by the kick \cite{HS}. Eventually, if
the kick is too strong, it destroys the soliton. In particular, for $%
\left\vert g_{1}\right\vert =1.2$ the destruction is observed at $v\geq 0.48$%
, see an example in Fig. \ref{fig6} for $v=0.6$.

\begin{figure}[tbp]
\begin{center}
{\includegraphics[width=7.cm,clip]{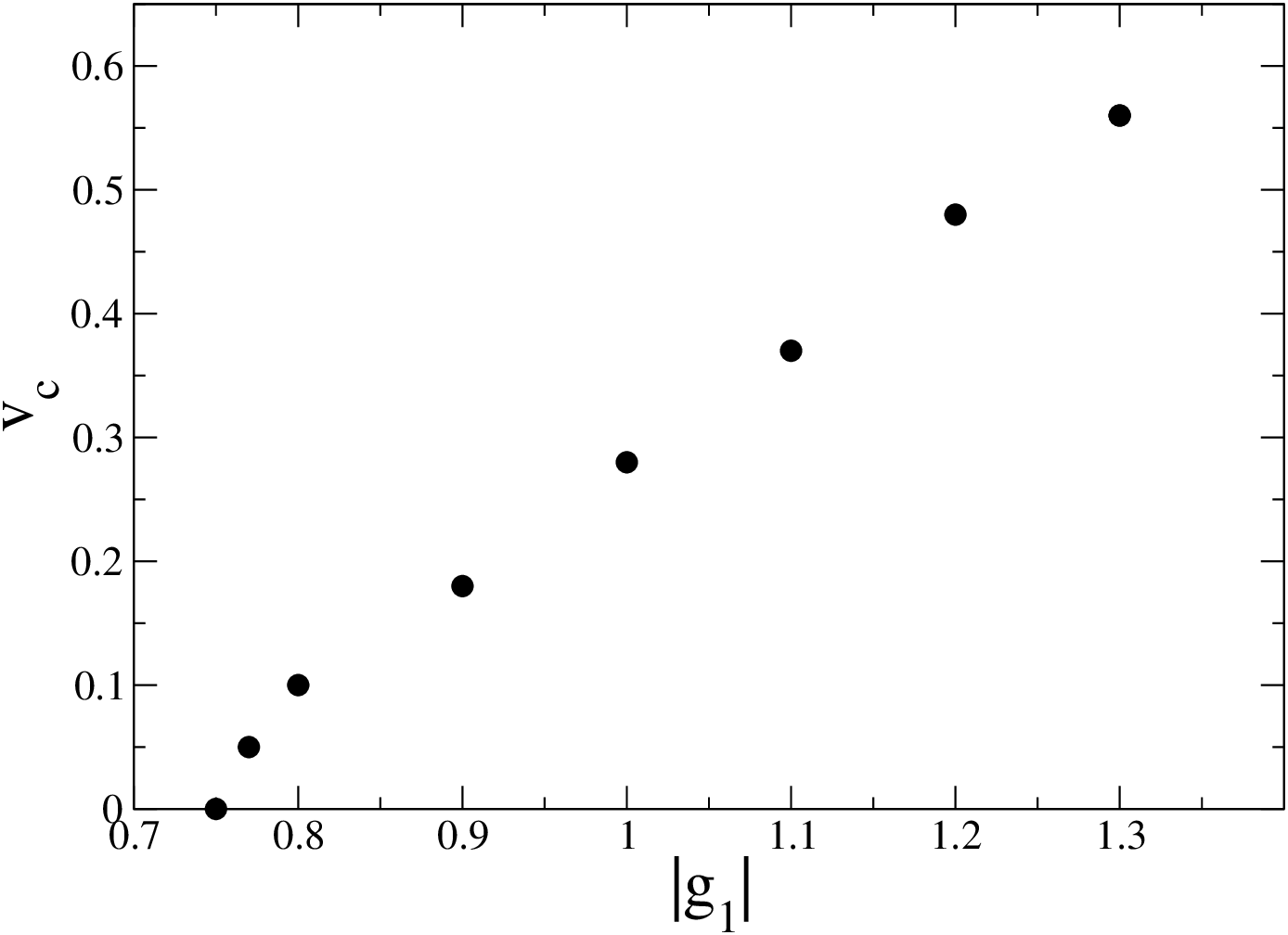}}
\end{center}
\caption{The critical velocity, $v_{c}$, for the destruction of the kicked
soliton versus the strength of the nonlinear lattice, $|g_{1}|$. Here, $%
g_{0}=0$ and $k=0.5$ are fixed, as before.}
\label{fig7}
\end{figure}

The critical velocity, $v_{c}$, at which the kicked soliton is destroyed, is
shown in Fig. \ref{fig7} as a function of the NL strength, $|g_{1}|$. The
figure features a linear growth of $v_{c}$ with NL strength $|g_{1}|$, at
sufficiently large values of $\left\vert g_{1}\right\vert $. This fact can
be explained by estimating the critical velocity as that at which the
respective kinetic energy of the kicked soliton, $(1/2)M_{\mathrm{sol}}v^{2}$%
, is equal to height $V_{\mathrm{PN}}$ of the effective Peierls-Nabarro
potential induced by the nonlinear (pseudo) potential. The mass of the
soliton, $M_{\mathrm{sol}}$, is proportional to its norm, which is $1$,
according to Eq. (\ref{N1D}). Further, the potential-energy density
corresponding to Eq. (\ref{1dnpse}) actually coincides with the potential
part of Lagrangian density (\ref{effective}), that should be evaluated with
the help of expression (\ref{sigma}) for $\sigma $. Then, a straightforward
consideration of Eqs. (\ref{effective}), (\ref{sigma}) and (\ref{1dnpse})
yields the following scaling relations in the limit of large $|g_{1}|$: $%
\left\vert f(0)\right\vert ^{2}\sim \sigma ^{-1}\sim W^{-1}\sim \left\vert
g_{1}\right\vert ^{-1}$, where $W$ is the axial size of the soliton, and,
eventually, $V_{\mathrm{PN}}\sim \left\vert g_{1}\right\vert ^{2}$. Thus,
the threshold condition, $(1/2)M_{\mathrm{sol}}v^{2}=V_{\mathrm{PN}}$,
explains the linear proportionality between $v_{c}$ and $\left\vert
g_{1}\right\vert $, which is observed in Fig. \ref{fig7} at large $%
\left\vert g_{1}\right\vert $. The vanishing of $v_{c}$ at $\left\vert
g_{1}\right\vert \approx 0.75$ in Fig. \ref{fig7} is a consequence of the
fact that the soliton with this value of $\left\vert g_{1}\right\vert $ lies
at the edge of the triangular area in Fig. \ref{fig2}, i.e., it does not
exist as a stable mode even without being kicked.

\begin{figure}[tbp]
\begin{center}
{\includegraphics[width=7.5cm,clip]{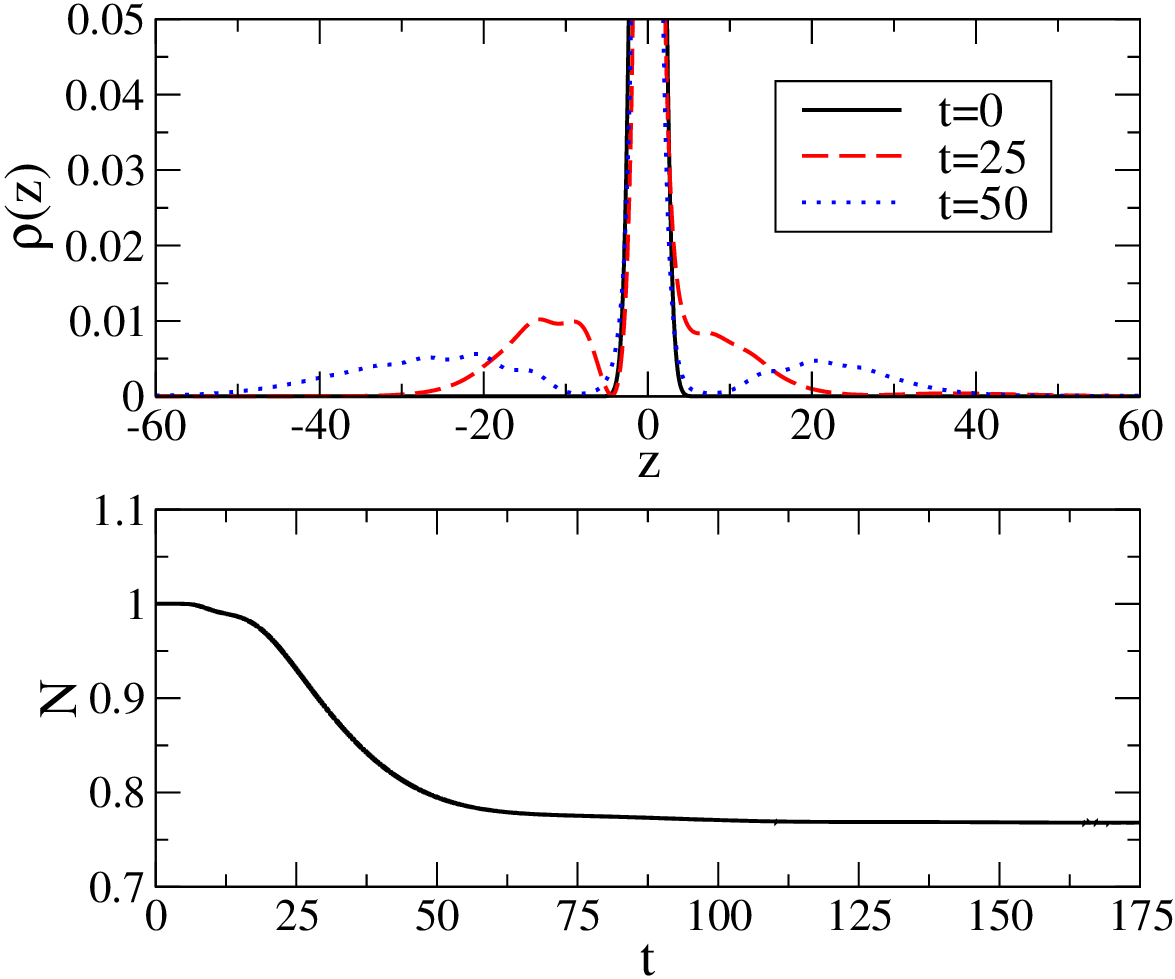}}
\end{center}
\caption{(Color online) Upper panel: The dynamics of the Gaussian wave
packet, kicked with initial velocity $v=0.4$, whose parameters belong to the
collapse domain in terms of Fig. \protect\ref{fig2}: $g_{1}=-1.5$, $k=0.5$,
and $g_{0}=0$. At $t=0$ (the initial condition shown by the solid line),
there is only the column centered at $z=0$, which represents the Gaussian of
axial width $1.7$. Lower panel: the norm of the wave function in interval $%
-15<x<15$, as a function of time.}
\label{fig8}
\end{figure}

The fact that the kick induces emission of radiation from the soliton may be
used to stabilize them in the collapse domain of Fig. \ref{fig2}. To this
end, we take, for example, the Gaussian wave packet with $g_{1}=-1.5$ and $%
k=0.5$, which falls into the region of the collapse. Real-time
simulations of Eq. (\ref{1dnpse}) demonstrate that, if the Gaussian
is kicked hard enough, it does not blow up, but rather forms a
stable soliton, in a
combination with the emission of radiation waves. This is shown in Fig. \ref%
{fig8} for initial velocity $v=0.4$. The initial configuration evades the
blowup because the emission of radiation reduces the norm of the remaining
soliton, pushing it beneath the collapse threshold, see the lower panel in
Fig. \ref{fig8}. This observation suggests that the relaxation of the
perturbed soliton via the emission of radiation proceeds faster than the
onset of the collapse, which attests to the robustness of the solitons.
Finally, as expected, if the kick is too hard (in the present case, this
means $v\geq 0.6$), it destroys the Gaussian wave packet, causing its
complete decay into radiation.

\section{Conclusions}

We have reported results for 3D matter-wave solitons supported by a
combination of the axial 1D NL (nonlinear lattice), which periodically
reverses the sign of the nonlinear interaction, and the tightly trapping
harmonic-oscillator potential acting in the transverse plane. The results
were obtained by means of two distinct approaches: The VA (variational
approximation), which was applied directly to the 3D Gross-Pitaevskii
equation, and the 1D NPSE, that was derived from the underlying 3D equation.
Previous works did not study the stabilization of 3D solitons by NLs. The
main result, produced by means of both methods in similar forms, is the
stability domain for solitons in the plane of the NL strength and
wavenumber. The usual 1D cubic Gross-Pitaevskii equation with the NL cannot
produce adequate results, as it does not give rise to the collapse, which is
the most important stability-limiting factor.

Another essential difference of the solitons produced by the NPSE with the
NL from their counterparts in the case of the cubic NL is that the solitons
are immobile in the framework of the NPSE: The kick applied to the soliton
either leaves it pinned, or, eventually, destroys it. The critical size of
the kick which destroys the soliton was found to be proportional to the
strength of the NL, provided that the strength is large enough; an
explanation for this dependence was proposed. On the other hand, the kick,
if applied to the wave packet created above the collapse threshold, may help
it to shed off the excess norm and thus stabilize itself against the
collapse. A related dynamical effect, which demonstrates the difference of
the NLs from linear lattices, is that wave packets relaxing into solitons
can emit small-amplitude waves, which freely propagate in the system.

A challenging extension of the analysis may be to develop it for the setting
with a 2D NL and the 1D trapping potential acting in the transverse
direction. The 2D version of the NPSE was developed previously, but in the
absence of the NL \cite{sala-npse,we2D}. In this case, one may expect the
existence of both fundamental and vortical 2D solitons.

\end{document}